# Unraveling BitTorrent's File Unavailability: Measurements, Analysis and Solution Exploration

Sebastian Kaune[1], Rubén Cuevas Rumín[2], Gareth Tyson[3], Andreas Mauthe[3],
Carmen Guerrero[2], and Ralf Steinmetz[1]

*Technische Universität Darmstadt* [1], *Universidad Carlos III de Madrid* [2], *Lancaster University* [3]

## Abstract

*BitTorrent suffers from one fundamental problem: the long-term availability of content. This occurs on a massive-scale with 38% of torrents becoming unavailable within the first month. In this paper we explore this problem by performing two large-scale measurement studies including 46K torrents and 29M users. The studies go significantly beyond any previous work by combining per-node, per-torrent and system-wide observations to ascertain the causes, characteristics and repercussions of file unavailability. The study confirms the conclusion from previous works that seeders have a significant impact on both performance and availability. However, we also present some crucial new findings: $(i)$ the presence of seeders is not the sole factor involved in file availability, $(ii)$ 23.5% of nodes that operate in seedless torrents can finish their downloads, and $(iii)$ BitTorrent availability is discontinuous, operating in cycles of temporary unavailability. Due to our new findings, we consider it is important to revisit the solution space; to this end, we perform large-scale trace-based simulations to explore the potential of two abstract approaches.*

## 1 Introduction

BitTorrent [3] has become a de-facto standard for scalable content distribution over the Internet. The reason for its success is its ability to efficiently leverage the uplink capacity of nodes whilst achieving high scalability during peak demands [10, 18]. This efficiency is largely attributable to BitTorrent's tit-for-tat mechanism, which encourages users to share their resources whilst downloading files.

Despite the success of BitTorrent, it still suffers from a significant problem: the *long term availability of content*. More specifically, content that is distributed using BitTorrent often becomes unavailable after a relatively short period of time. For example, [7] found that the available lifespan of most torrents is between 30-300 hours whilst 10% of all users fail to successfully download their desired content.

A file can be considered unavailable if one or more of its data pieces are inaccessible to users wishing to download it. The most intuitive reason for this occurrence is that previously successful users in possession of the entire file (seeders) have left the system leaving only users that possess a subset of the file (leechers). Subsequently, unavailability occurs when this subset cannot collectively rebuild the complete file with their remaining pieces. Previous research (such as [8][7][13]) has promoted the importance of seeders in regard to availability and concluded that a seedless torrent is unable to reconstruct the file. However, this conclusion is challenged by the observation that some torrents continue to effectively serve files despite lacking any seeders.

In this paper, we devote our attention to understanding and characterizing BitTorrent's file unavailability problem. We strive to discover the scale, causes and repercussions of the problem alongside investigating the possible solution space. To achieve this we have performed two large-scale measurement studies; the first investigates BitTorrent on a macroscopic level by periodically probing over 46K torrents to ascertain their high level characteristics, such as swarm size and seeder/leecher ratio. Whilst, the second study investigates BitTorrent on a microscopic level by contacting over 700,000 individual peers in 832 torrents to discover relevant properties such as their download rates and piece availability. To the best of the authors' knowledge, this is the largest dataset in terms of size and collected information used to investigate file availability in BitTorrent. This allow us to extend previous works to obtain far more accurate results; through this we make a number of interesting findings,

- In 86% of cases, leechers are unable to reconstruct files in the absence of seeders. However, in 14% of cases, leechers can reconstruct the file without any seeders present. We therefore discover that seeders are not the sole factor involved in BitTorrent's unavailability problem. Such torrents achieve this through the possession of large and stable populations as well as high aggregate download rates that enable leechers to quickly replicate rare chunks.

- In 64% of torrents, unavailability is *not* immutable and, instead, occurs in cyclic periods followed by reoccurring availability. This is due to old seeders returning to swarms where they previously participated in.

- The combination of the two previous observations results in 23.5% of users affected by a lack of seeders actually being able to complete their downloads.

- Users often become frustrated with unavailable torrents that exhibit poor download rates. We observe a chain reaction in which such users abort their downloads thereby exacerbating unavailability, resulting in further abortions.

These new findings make it crucial to revisit the solution-space to investigate behaviour under the new, accurate workload defined by our large scale dataset. As such, we perform trace-based simulations looking at both traditional single-torrent and cross-torrent mechanisms approaches to solving the file unavailability problem; our primary results are,

- Single-torrent incentive mechanisms must encourage users to increase the average seeding time to 10 times more than the current average to achieve 99% availability.

- Cross-torrent incentive mechanisms can easily achieve 99% availability but with a performance decrease of 22% for 56% of the users.

The rest of the paper is structured as follows; Section 2 provides related work. Section 3 then details the problem and our measurement methodology. Following this, Section 4 characterises the causes and impact of unavailability. We discover a primary cause is a lack of seeders and therefore Section 5 investigates seedless states in BitTorrent. Next, we utilise our measurement study data to explore the potential solution space with trace-based simulations in Section 6. Finally, we conclude the paper in Section 7.

## 2 Related Work

**BitTorrent Measurements**: BitTorrent measurement studies can be classified into two different groups. The first type uses log traces from trackers [10, 8, 7, 1] whereas the second type relies on crawling techniques to retrieve the information from the system [18, 19, 21, 13, 17]. The first type of measurements is less intrusive since they do not actively interfere with the system. However, they are often problematic to obtain since they require the agreement from content providers. The crawling techniques, on the other hand, can be divided into two categories. In its simplest form, a crawler exploits the BitTorrent protocol to periodically request the IP addresses of the clients participating in the torrent from the tracker [21]. This makes it possible to study the demographics and dynamics of the torrents under analysis. This is what we name *macroscopic crawling*. More sophisticated crawlers also contact the clients and retrieve detailed information such as the client ID and their piece bitmap. We name this *microscopic crawling*. Although the microscopic crawling gives more detailed information, it is noticeably less scalable and only allows a few thousand torrents to be studied in parallel [18, 19]. Each approach is effective for addressing particular needs; however, these have not yet been combined to investigate BitTorrent in a holistic way.

**BitTorrent's File Availability Analysis**: There are only a few works investigating availability issues in BitTorrent systems [7, 15, 13]. Neglia et al. mainly study the tracker/DHT availability of 22,000 torrents obtained from two torrent indexing sites [15]. Guo et al. [7] extended this, to model the lifespan of torrents by analyzing a limited number of tracker traces from [15]; it was found that most torrents are short-lived because of an exponentially decreasing peer arrival rate. This model starts from the basis that content is unavailable when there are no seeders present in the swarm. This is, so far, an unverified hypothesis that is important to investigate. Similarly, Menasche et al. also use this hypothesis to investigate the availability of seeders in 45,000 torrents obtained from the Mininova website [13], finding that 40% of swarms lack seeders for more than 15 days in the first month after the torrent's birth.

**Improving BitTorrent File Availability**: Surprisingly, little research work has been performed into addressing the file availability in BitTorrent [8, 13, 20]. The most recent work improves file availability problem in BitTorrent by file bundling to enlarge the online times of the users. Using a queuing theoretic model and controlled experiments on PlanetLab, the authors show that this approach can reduce waiting-time for peers in torrents with highly unavailable seeders. However, their results consider that peers arrive in a constant Poisson process which is a strong assumption given the measurement results presented in [18, 8] and also in this paper.

Guo et al. were the first to propose intriguing ideas and results for *cross-torrent collaboration*. Amongst other things, the authors sketch an abstract mechanism for instant inter-torrent collaboration; following this they also evaluate the principles. Yang et al. propose a variation of these ideas by designing a cross-torrent tit-for-tat strategy that assumes repeated interactions of the users. However, this method suffers because as Piatek et al. show through extensive measurements, 91.5% of peer pairs that occur in a single swarm will never meet again at any later point in time [17]. Pi-atek et al. subsequently propose an alternative protocol that enables long-term incentives in BitTorrent with the aid of one-hop intermediaries.

## 3 Problem Background and Methodology

### 3.1 Defining File Availability

To study and understand the availability of files in BitTorrent we first present a simple model. Let's assume that we have a torrent $T$, formed by $N$ nodes, managing the download of a file composed by $P$ pieces. Thus, we can define the vector $V_i = [V_{i1}, V_{i2}, ..., V_{iP}]$ that contains the information about the pieces stored by peer $i$: $V_{ij} = 1$ if node $i$ has the piece $j$; $V_{ij} = 0$ if node $i$ does not have piece $j$. $V_i$ is typically known as the *bitfield* of node $i$.

We define the *Percentage of Available Pieces* of torrent $T$ at a time instant $t$ as

$$U(T) = \frac{\sum_{j=1}^{P} OR(V_{ij})}{P}. \quad (1)$$

Where $OR(V_{ij})$ represents the logical *OR*-operation over the piece $j$ across all the nodes in the torrent $T$.

### 3.2 The Circumstances of Unavailability

It is important to understand in which circumstances a file becomes unavailable, based on our definition. A file is considered unavailable if at least one of its pieces is not accessible within a swarm. This situation arises if there are no peers in the swarm that possess a given piece or, alternatively, if the peer(s) that possess the piece are inaccessible (e.g. due to firewalls, NAT or overlay graph disconnection). It is intuitive to consider the former as a far more likely circumstance (e.g. most BitTorrent clients implement techniques such as NAT traversal [14]. Moreover, they include neighbors discovery techniques such as the Peer Exchange Protocol -PEX- and periodical tracker polling that prevent graph disconnection). Therefore, given this assumption, a file can be considered available if ($i$) there is at least one seeder or ($ii$) there is no seeder but the bitfields of the leechers collectively fit the condition $U(T) = 1$. Without detailed analysis, we can therefore currently state that:

- With an accessible seeder, a file *is* available
- Without an accessible seeder, a file *may* be available

This paper uses these two observations as a starting point to investigate unavailability in BitTorrent. In the following sections, we denote time periods in a torrent's lifecycle in which no seeder is online as a *seedless state*. To this end, the file is *unavailable* if torrent $T$ is in seedless state and $U(T) < 1$.

### 3.3 Measurement Methodology

To study the unavailability problem and specifically the seeders' role in it, we have performed two large-scale measurement studies using microscopic and macroscopic crawling. To the best of our knowledge, this paper is the first to combine both microscopic and macroscopic crawling techniques to better understand BitTorrent (specifically BitTorrent's file availability).

*Microscopic Crawling*: To truly understand unavailability in BitTorrent, it is necessary to be able to view the microscopic characteristics of any given swarm, e.g. piece distribution or nodes' download rates. Without this, one can only get a rough estimation of availability using metrics such as the number of seeders. The information regarding the behaviour of individual peers provides the necessary data to make new, more accurate findings. To gain this information we developed and deployed a distributed BitTorrent crawler that can investigate swarms on a microscopic level, using 20 nodes in the Emulab testbed [5].

The crawler operated from July 18, 2009 to July 29, 2009 (`micros-1`) and then again from August 19, 2009 to September 5, 2009 (`micros-2`). To discover all the online users in a torrent it periodically contacted the torrent's tracker as well as using the Peer Exchange Protocol (PEX). From every peer, every 10 minutes it requested their piece bitmap to discover the real-time distribution of pieces. For the `micros-1` study, the crawler followed 255 torrents appearing on Mininova[1] after the first measurement hour; in these torrents, we observed 246,750 users. The `micros-2` dataset contains information from 577 torrents and 531,089 users.

*Macroscopic Crawling*: The microscopic measurements provide detailed insight into the distribution of pieces and download rates within the swarm, as well as between different peers. However, due to scalability issues it is difficult to perform such detailed measurements on a very large-scale (e.g. several thousand torrents). To complement these results we therefore also implemented a higher level crawler that followed every torrent published on the Mininova website after December 09, 2008 for a period of 38 days. This crawler periodically requested, from multiple sites in Europe, tracker information regarding each torrent's number of seeders and leechers alongside the members' ip addresses (we were able to systematically collect 98% of all the ip addresses from within the swarms). This study allowed us to gain an extremely large number of measurements regarding details such as peer arrival patterns, seeder/leecher ratios and torrent sizes. This information can subsequently be correlated with our smaller-scale microscopic measurements to derive such things as the scale of seedless states and the causes for seedless states occurring. Our final macro-scopic dataset consisted of reports from 46,227 torrents and 29,066,139 users.

---
[1]The largest BitTorrent Community based on Alexa Ranking.

## 4 Characterising Unavailability: Causes and Impact

In this section, we first investigate the role that seeders play in file unavailability. Following this we study the exceptions and variations we discovered. Lastly, we then investigate the real-time impact that a lack of seeders has on client performance and their subsequent reactions that can be observed.

### 4.1 Investigating the Role of Seeders in File Unavailability

It is intuitive to think that U(T) < 1 in a torrent without any seeder (that is, leechers are unable to reconstruct the file). However, this is, so far, an unverified assumption that must be investigated (and quantified). To ascertain this, we inspect the *(i)* nodes' bitfield and *(ii)* nodes' download rates in all the torrents of our microscopic traces affected by seedless states.

#### 4.1.1 Bitfield Analysis

We have collected every nodes' bitfields for all the torrents in our microscopic measurements as they have evolved over time. For each torrent we have computed $U(T)$ periodically every 10 minutes during any period a torrent is without any seeders (i.e. it is in a seedless state). This allows us to ascertain whether a full copy of the file exists in the torrent at any given time. Fig. 1 shows the CDF of $max(U(T))$ observed in the seedless state for each torrent that we studied. From this data, we can extract two pieces of information; first, in the majority of cases (86%) our hypothesis is confirmed and the contacted leechers are unable to collectively reconstruct the file once a seeder has left (i.e. $max(U(T)) < 1$). Clearly, this means that seeders *do* have a significant impact on the availability of files in BitTorrent. Importantly, however, we also find that a notable proportion of torrents (14%) actually remain available even without a seeder. Collectively, this makes up 24% of all leechers that operate in seedless swarms. This is a crucial finding that has not been observed before; it is therefore in contrast with previous models [13, 8] that consider all seedless torrents to be unavailable.

#### 4.1.2 Download Rate Analysis

A limitation of the bitfield analysis is that not all nodes are accessible due to NATs. To address this, we also inspect the aggregate torrent download rates. Through this, we can infer that a file is unavailable when the download rate of all the peers participating in a specific torrent drops close to 0 KBps. From this we can derive that the node cannot find any new pieces to download.

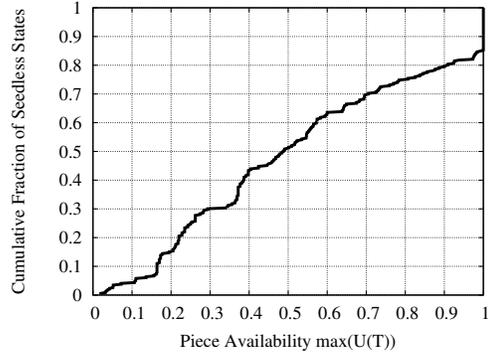

**Figure 1. Piece availability in torrents affected by seedless states.**

To highlight our findings, we first inspect a representative torrent from our microscopic trace[2], shown in Fig. 2. The figure shows the median instant download rate of the online leechers over time, sampled every 10 minutes. It also plots the number of seeders and leechers, as well as the number of copies of the least replicated piece. Note that when the number of seeders becomes 0, the torrent enters a seedless state.

The torrent can be observed to enter a seedless state after the middle of day 3, remaining in this state for roughly two days. When the final seed departs the download rate of the leechers drops to approximately 0-3 KBps after only a few minutes. This also coincides with the number of least replicated pieces dropping to zero. It can therefore be confidently inferred that the file is, indeed, unavailable during this period due to the departure of the last seeder.

Interestingly, it can also be seen that the torrent becomes available again during day 5. As the seeders return, the download rate increases and the file becomes available again. In contrast to past assumptions, it is therefore evident that unavailability is not continuous. This important phenomenon will be investigated further in Section 5.3.

The above analysis has inspected a representative torrent. To validate its widespread applicability we also look at the download rate degradation in all torrents. To achieve this, we have taken all the users that have been affected by a seedless state and separated their downloading time into two periods: ($i$) periods in which they have suffered from a seedless state and ($ii$) periods in which they have not. Fig. 3 presents the download rate distribution for both periods. First, we can observe that the download rate in a non-seedless state is much higher than in a seedless state. 80-85% of the nodes experience an average download rate lower than 1 KBps when in a seedless torrent, indicating that the peers cannot locate any required pieces and the file is, indeed, unavailable. Second, however, we also observe that 15-20% of users, in fact, maintain a reasonable level of performance even without any seeders. This can be attributed to two reasons: $(i)$ the aforementioned 14% of torrents are capable of reconstructing their file without a seeder at an average rate of 21.3 KBps; and $(ii)$ newly joined peers can download the subset of available pieces at an effective rate. This can be observed in the representative torrent (cf. Fig. 2): between days 4 and 5 there is a peak in the number of leechers which results in a short peak in the download rate as new comers download the available pieces.

---
[2]We have observed the same behaviour in most of the torrents affected by seedless states.

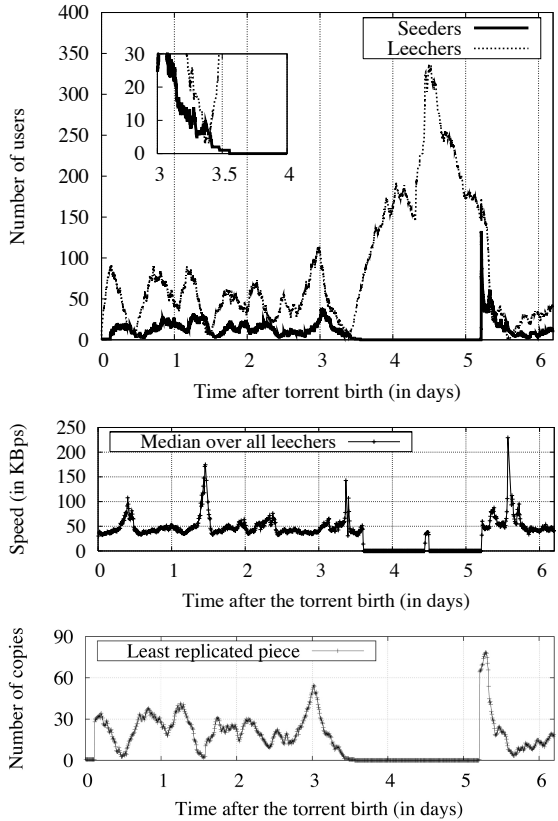

**Figure 2. Snapshot from a torrent in our microscopic trace.**

### 4.2 Investigating the Causes of Swarm Resilience

The previous section has identified a notable percentage (14%) of torrents that can maintain availability even without any seeders; this represents 24% of all leechers that encounter seedless states.

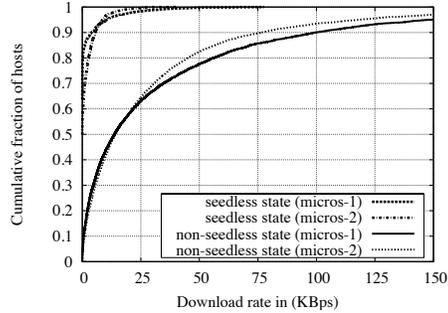

**Figure 3. Interval download rates for nodes affected by the lack of seeders.**

To investigate this, we separate torrents into those that survive in the absence of seeders (*resilient torrents*) and those that do not (*susceptible torrents*). We then investigate quantitative properties of these two groups to ascertain how they differ at various points in their lifecycles. All of the identified metrics have been calculated for each group (across all member torrents) every 10 minutes using information from the microscopic traces and the tracker reports. These values have then been averaged together over each time period investigated.

Table 1 gives an overview of all metrics used in this analysis. We calculate these over two time periods: the beginning of the torrents' lifecycle and just before the last seeder goes offline. Although not included in the table, we also investigated the effects of file size and content type without ascertaining any correlation. Most metrics are straightforward, however, two require some explanation: Distribution Entropy ($E(T)$) and the Churn Factor ($CF$).

The $E(T)$ investigates the distribution of pieces within the swarm; this is to investigate whether torrents that can survive achieve a superior distribution of pieces. We therefore characterize the distribution entropy in a torrent $T$ at a given time $t$ by introducing the following *Entropy Index*:

$$E(T) = \frac{\left(\sum_{j=1}^{P}\sum_{i=1}^{N} V_{ij}\right)^2}{P \cdot \sum_{j=1}^{P}\sum_{i=1}^{N}(V_{ij})^2} \quad (2)$$

Recall that $N$ defines the number of nodes in the swarm, $P$ is the number of pieces a file is composed of and $V_i$ is the bitfield of node $i$. This index is similar to Jain's Fairness Index [11] and achieves a value of 1 if all pieces are equally distributed among the peers.

The Churn Factor $CF$ investigates whether torrents that can survive have more stable populations. This factor is defined by $N_{disc}/N_{all}$ where $N_{disc}$ is the number of users that have left the swarm during a given time period ($t$) and $N_{all}$ is the total number of users observed during this same period. A factor of 0 indicates that no user disconnected within $t$; by default $t = 10\ mins$.

| Metric | Time before seedless state | | | | Time after torrent's birth | | | |
|---|---|---|---|---|---|---|---|---|
| | 1 hour | | 6 hours | | 6 hours | | 24 hours | |
| | Resilient | Susceptible | Resilient | Susceptible | Resilient | Susceptible | Resilient | Susceptible |
| Swarm speed (in KBps) | 58.83 | 23.88 | 68.58 | 24.99 | 95.72 | 53.50 | 62.99 | 44.82 |
| Seeder/Leecher ratio | 0.15 | 0.03 | 0.14 | 0.04 | 0.19 | 0.19 | 0.32 | 0.43 |
| Firewalled/NATed peers (in %) | 70.86 | 61.09 | 62.30 | 60.67 | 51.30 | 56.05 | 54.94 | 58.44 |
| Distribution Entropy $E(T)$ | 0.93 | 0.94 | 0.92 | 0.93 | 0.91 | 0.90 | 0.92 | 0.91 |
| Least replicated piece (# of copies) | 9.21 | 1.61 | 8.38 | 2.82 | 15.34 | 14.21 | 21.33 | 23.55 |
| Churn factor $CF$ | 0.03 | 0.21 | 0.08 | 0.15 | 0.07 | 0.11 | 0.08 | 0.07 |
| Online leechers | 281.34 | 111.61 | 250.48 | 101.55 | 134.05 | 82.71 | 163.12 | 89.55 |
| Online seeders | 5.28 | 1.51 | 6.11 | 1.80 | 12.05 | 11.25 | 18.17 | 21.34 |

**Table 1. Characteristics of resilient torrents (those that maintain availability in seedless state) and susceptible torrents (those that cannot reconstruct the file).**

From the data in Table 1, we can make the following important observations,

- *Torrent Popularity:* From the beginning, resilient torrents exhibit higher leecher population sizes. Larger torrents possess an increased probability of replicating rare pieces before the loss of seeders.

- *Low Churn Factor:* High churn in small torrents creates a greater risk of losing vital pieces; if this coincides with the loss of a seeder then it becomes impossible to recover these pieces again until a seeder returns. Resilient torrents have significantly lower churn factors than susceptible torrents.

- *Seeder/Leecher Ratio:* Resilient torrents exhibit a higher seeder/leecher ratio and, as a derivative of this, experience download rates that over twice as high as susceptible torrents. This superior performance is highly beneficial for the survival of piece replicas as it allows the quick duplication of rare pieces. Before seedless state occurring, resilient torrents therefore have many more replicas of the rarest piece when compared to susceptible torrents.

In summary, these results show that swarm resilience is a product of large, stable populations that can achieve higher download rates due to beneficial seeder/leecher ratios. The combination of these factors results in rarest piece replication rates that are over 5 times greater than their susceptible counterparts. This makes such swarms highly resilient to the loss of any seeders. Importantly, it also can be concluded that unavailability cannot be addressed by modifying any of BitTorrent's algorithms (e.g. piece selection) but, instead, must be solved by incentivising users to modify their behaviour. This is exemplified by the lack of any correlation between resilience and distribution entropy.

### 4.3 Effects and Trends of Seed Departure

Despite a notable percentage of torrents surviving without seeders, it is evident that the loss of all seeders can often result in unavailability. This section now investigates the effects that this has on both individual users and wider system performance. Three stages can be identified which we now discuss.

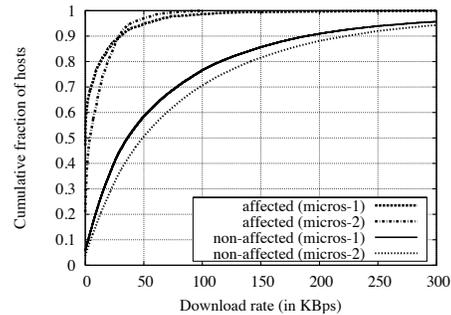

**Figure 4. Comparison of download performance between peers affected and not affected by seedless states.**

The first repercussion of the loss of seeders in susceptible torrents is a significant and rapid drop in download rates. To extend the earlier analysis, we now compare the average download rate of users that suffer from a seedless state at some point during their download against the average download rate of users that always find content available. We first categorise users into two groups: affected vs. non-affected. The first group of users consists of leechers that are (at some point) affected by a lack of seeders. The non-affected users, on the other hand, have at least one seeder available during their entire download. Fig. 4 gives the download rate distribution of both user groups as obtained from the two microscopic crawlings. Whereas the median download rate for the non-affected users is 36 KBps in `micros-1` and 48 KBps in `micros-2`, the performance for peers attempting to download unavailable content is only 0.06 KBps and 3.8 KBps, respectively.

The second observable stage is a direct derivative of the decrease in download performance. Specifically, we observe a large increase in download abortions. To study this we examine the session times in our microscopic traces. We observe that 89% of users affected by file unavailability (i.e. participating in *susceptible* torrents) abort their downloads due to the bad performance. Sadly, this is an unnecessary action as we have found that seeders often return, making files available again. In contrast to these results, users operating in resilient torrents only have an abortion rate of 34.47%. Although this seems initially high, we also find that many users operating in other torrents that do not suffer from unavailability also abort their downloads. On closer inspection, these 'unnecessary' abortions occur in torrents that have particularly low download rates that are under a third of the average.

The third stage in this process is the worrying emergence of a chain reaction. We find that as the number of abortions increase, the number of available chunks decrease. This results in an exacerbation of the torrent's unavailability and a further drop in download rates for those trying to access the remaining chunks. As other users witness this trend, they too abort their downloads. This process results in fewer users becoming seeders and therefore greater unavailability and more abortions. Frequently, the above two repercussions of unavailability and the creation of this chain reaction often spells the end for a torrent.

From these findings we derive that users are highly sensitive to their perceived instant quality of service and therefore any solutions must maintain an acceptable download rate whilst also improving file availability.

## 5 Characterising Seedless States

The previous section has validated and quantified the importance of seeders in regard to file availability in BitTorrent and discussed under which circumstances a file of a seedless torrent becomes unavailable. It has been found that in the majority of torrents (86%), the loss of all seeders results in unavailability. In this section we therefore investigate the behaviour of seeders and characterise the nature of seedless states using our large scale dataset. We first look at the frequency of seedless states in BitTorrent. Following this we investigate the causes of seedless states before, finally, investigating the issue of why torrents can become revived again after extended periods of unavailability.

### 5.1 How Prevalent are Seedless States?

To quantify how prevalent seedless states are in BitTorrent, we ask the following question: how *many torrents* and to *what extent* are torrents affected by seedless states?. To answer this, we use the logs from our macroscopic trace that give us a large-scale view on the system comprising of 46k torrents.

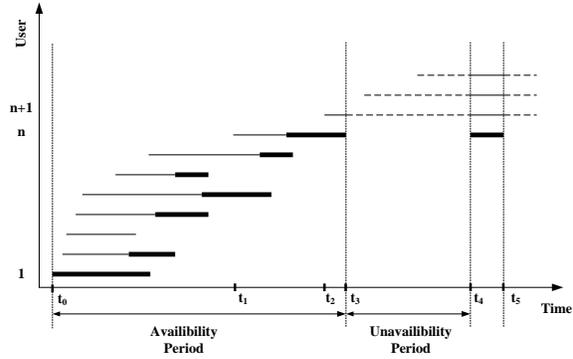

**Figure 5. Illustration of a seedless state.**

The measurements show that more than 38% of torrents (17,568 out of 46,227) lose their seeders within the first month, out of which 72% lack seeders after only 5 days. Similarly, we find that more than 45% of the torrents suffer from a lack of seeders for half of their monitoring time. To exemplify the scale of this, in 50% of the torrents observed for periods longer than 30 days, no seeder was available for more than 16 days.

Finally, in our study, more than 9.68 million users (33% of all users seen) participated in torrents with highly unavailable seeders suggesting that this is not only a long tail problem. Out of these users, more than 1.59 million were directly affected by seedless states.

### 5.2 Why do Seedless States Occur?

Since seedless states are highly prevalent in real swarms, an intuitive question is: *why do they occur in the wild?* In this section, we first identify and then further investigate the influencing factors responsible for triggering seedless states.

#### 5.2.1 Identifying Influencing Factors

There are two main factors that directly influence the existence of seedless states: ($i$) the session time of seeders and ($ii$) the inter-arrival rate of the users. To illustrate the influencing factors, we use a simple example shown in Fig. 5. In this figure, each horizontal line represents the lifetime of a user; these users can either be in a leecher state (thin lines) or a seeding state (thick lines).

It seems straightforward that the longer a seeder serves content, the more leechers are able to finish their downloads. Unfortunately, (as demonstrated later on) the seeding time is typically quite short contributing significantly to the frequency and length of seedless states.

Let's now assume that user $n$ is the last available seeder in our example torrent and none of the previous seeders return to the torrent. In this case, a seedless state occurs when the time required for leechers to download the file exceeds the online time of the last seeder. For example, Fig. 5 shows that after the last available seeder leaves the swarm at time $t_3$, none of the remaining leechers were able to finish the download. If we focus on the $n$-th node and its subsequent successor in the torrent ($n+1$-th), the inter-arrival time between both users is given by $\tau_{n+1}(= t_2 - t_1)$ whereas the seeding time of node $n$ is given by $\mu_n$. Assume that both users $n$ and $n+1$ download a file of size $F_s$ with rate $D_n$ and $D_{n+1}$ respectively. Thus, the swarm enters a seedless state when Eq. 3 is fulfilled.

$$D_n F_s + \mu_n < \tau_{n+1} + D_{n+1} F_s \qquad (3)$$

To simplify the analysis, we assume that $D_n = D_{n+1}$[3]. In this case, the seedless state is reached if the inter-arrival time is larger than the seeding time.

To summarise, seeding times as well as inter-arrival times play an important role in the generation of seedless states and subsequently in the long-term availability of content. Since both parameters are not directly correlated, we individually analyse both of them in the following.

### 5.2.2 Arrival Behaviour of Users

The first behavioural characteristic that is paramount to seedless state generation is the inter-arrival times of users. In this regard, intuitive questions are: *(i)* what inter-arrival times do we expect in reality and *(ii)* how do inter-arrival times evolve over time?

By analysing a few hundred torrents in a small community, previous work [7] has shown that user inter-arrival times are exponentially increasing. Our goal is to generalize this finding for 'open' communities such as Mininova.org that are orders of magnitude larger. For our analysis, we use similar techniques as applied in [7]. We consider all torrents in our macroscopic trace. We use linear regression to fit the logarithm of the complementary[4] of the number of node arrivals of each torrent along time. Let $X_t$ denote the complementary number of node arrivals at time epoch $t$ and $Y_t$ be the fitting result. We define the relative deviation of the actual node arrivals over an ideally exponentially increasing function by $\frac{log X_i - log Y_i}{log X_i}$. Thus, a relative deviation of 0% indicates that both curves overlap. Fig. 6 shows the deviation for each torrent of our macroscopic trace. The x-axis depicts the torrents ordered by ascending population size while the y-axis shows the relative deviation. For most of the torrents, the relative deviation is less than 10% whereas the deviation tends to decrease with increasing torrent popularity. Altogether, the average relative deviation of all torrents is 4.8%. Therefore, we conclude that the inter-arrival time of the nodes exponentially increases with time.

---
[3]Our microscopic measurements show that the download rate of users that finish downloads ($D_n$ in the example) is higher than the download rate of those that do not ($D_{n+1}$) validating our assumption.

[4]We use the complementary number of node arrivals to avoid domains in which the logarithm is undefined, e.g., epochs with no peer arrivals.

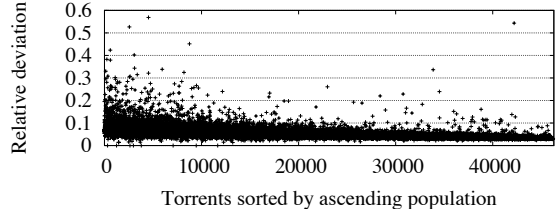

**Figure 6. Deviation from linear regression.**

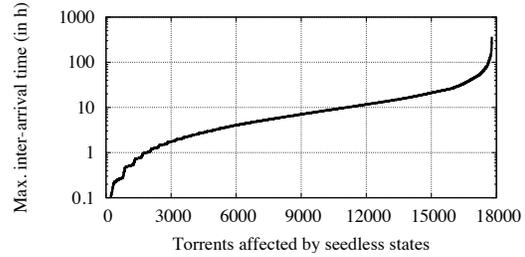

**Figure 7. Maximum inter-arrival times of torrents with highly unavailable seeders.**

Notably, we observed especially high inter-arrival times in torrents affected by seedless states; this is in line with our analysis in the previous section. For instance, Fig. 7 plots the maximum inter-arrival time observed in these torrents with unavailable seeders. More than 45% of the torrents exhibit inter-arrival times far beyond 10 hours.

### 5.2.3 Seeding Times of Users

The second behavioural characteristic that is paramount to the creation of seedless states is the seeding time of a node, i.e. how long seeds stay online for. As already shown in our example torrent (cf. Fig. 5), to maintain file availability it is necessary for seeders to remain online for long enough for new seeds to be generated. Fig. 8 shows the cumulative distribution of the seeding times of the nodes obtained from the two microscopic measurements. It can be seen that seeding times are generally short-lasting with 75% of the seeders staying online for less than 4 hours. When this data is compared to the inter-arrival time of users it can be identified that the current seeding times in BitTorrent are not sufficient to avoid seedless states, thus preventing to achieve long-term file availability in BitTorrent.

### 5.3 How long are Seedless States?

The representative snapshot presented in Fig 2 has highlighted that torrents can become available again after a ex-tended periods of unavailability. In this section we validate (using both the macroscopic and microscopic datasets) that file unavailability is, in fact, discontinuous with reoccurring periods of temporary availability. Through our measurement studies we can state that this occurs because seeders often return to swarms that they have previously participated in. This allows the 11% of users that choose to remain online during periods of unavailability (i.e. in *susceptible* torrents) to eventually complete their downloads. Alongside the existence of *resilient* torrents, this means that 23.5% of all leechers affected by seedless states can actually still gain access to the file.

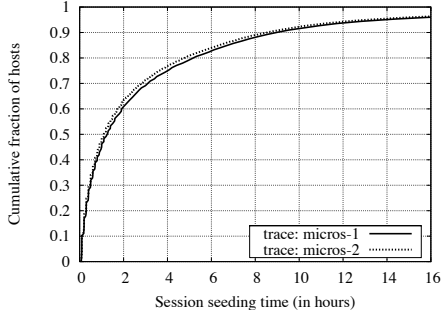

**Figure 8. Seeding time distribution.**

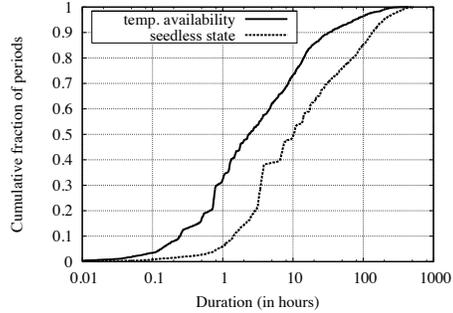

**Figure 9. Length of cycles of temporary availability and unavailability periods.**

The reoccurrence of seeders happens in over 64% of torrents that suffer from seedless states in our macroscopic study. To investigate this, Fig. 9 shows the CDFs of both the duration of seedless states as well as the duration of the subsequent periods in which content becomes available again, computed over all torrents exhibiting this phenomenon. Note that the x-axis is in log scale. It can be observed that seedless periods are typically long-lasting with an average of 43.19 hours whereas the subsequent availability periods only last 12.56 hours on average.

The primary reason for the (seemingly) altruistic return of seeders is likely to be the default settings of many BitTorrent clients (e.g. Vuze, $\mu$Torrent) that automatically rejoin torrents at their start-up even after a user has completed a file download. Unfortunately, BitTorrent users do not have permanent identifiers and thus we cannot make quantitative statements on exactly how many unique seeders rejoin a swarm and over what time period. However, the length of the seedless periods as depicted in Fig. 9 offers a conservative bound for the inter-seeding time distribution of such users. This is obviously very coarse grained and therefore we expect the inter-seeding times actually to be higher.

The reoccurrence of seeders is obviously in contrast to previous work that has assumed unavailability is continuous and immutable. To investigate the impact that this finding has on previous work, we briefly look at the relative deviation that the assumption has when compared against our dataset. We define the relative deviation as,

$$\text{relative deviation} = \frac{\text{measured avail. time} - \text{assumed avail. time}}{\text{assumed avail. time}} \quad (4)$$

We find that in approximately 35 % of the torrents in our macroscopic dataset, the assumption works well. In these torrents, we did not observe any temporary availability period after the torrent first enters a seedless state. However, in 50% of the torrents, the content is actually available for at least twice as that assumed when considering immutable unavailability.

## 6 Improving File Availability

The previous sections have outlined the file availability problem and highlighted the significant impact that seeders have on this. We therefore deduce that a solution must find some way to encourage users to provide content even after they have obtained it themselves i.e. to prevent seeders from leaving torrents. To do this we highlight two possible approaches: single-torrent and cross-torrent mechanisms. The principles of these two approaches are first abstractly outlined to show how each might improve seeding times. Following this, the two approaches are evaluated considering the key findings described in previous section (e.g. reoccurrence of seeders). For this purpose, we run trace-based simulations using the workload from our large scale dataset.

### 6.1 Potential Solution Approaches for Extending Seeding Times

This section briefly outlines the two generic approaches that can be taken for improving seeding in BitTorrent. The first is using traditional single-torrent principles whilst the second exploits the concept of cross-torrent collaboration (originally outlined in [7]). Note that we do not offer concrete implementational details; instead, we provide a brief outline of the principles behind each mechanism.

### 6.1.1 Single-Torrent Solution

A single-torrent solution involves incentivising users to remain within a torrent to seed, based on certain properties related to that individual torrent. As of yet we do not know of any successful mechanisms to achieve this due to the difficulty of enforcing incentives once a peer has already obtained the file which it desires. We therefore consider a simple framework of encrypted pieces that may work. Such a solution would involve encrypting the file before it is distributed within the swarm. The tracker would be responsible for managing this encryption and, as such, would be the source of the keys. Subsequently, once a peer has downloaded the file it would be required to remain seeding for a length of time determined by the tracker before the encryption keys are released to it.

### 6.1.2 Cross-Torrent Solution

A cross-torrent solution involves incentivising users to cooperate with the *system* as opposed to individual torrents. This approach is motivated by observations from our macroscopic trace that shows 51% of the users join multiple torrents (4.98 on average). We have further found that seeders frequently rejoin swarms after they have left, therefore providing conclusive evidence that the same peers rejoin the BitTorrent system multiple times whilst still possessing their previously downloaded files. To highlight the principles of a cross-torrent solution, imagine a user who joins torrent $X$ at some point in time and completes the download; this user may very well join another torrent $Y$ at a later point in time. When the node comes online again to download torrent $Y$ it could then theoretically persists as a *replica* for torrent $X$.

The incentives behind this could be managed in a number of ways (e.g. [17]). The following example highlights how the system could work using persistent contribution histories. Through this approach, the system would maintain a history of the contributions made by each user (agnostic to which torrent the contribution is made). Subsequently, peers would show preference to piece requests from users with higher contribution ratios. This would therefore replace BitTorrent's current rate-based tit-for-tat mechanism so that incentives were based on the entire system as opposed to individual users and torrents.

## 6.2 Experimental Methodology

To evaluate the two possible solutions approaches, the BitTorrent simulator of Bharambe et al. [2] is used and extended to enable the simulation of multiple torrents existing in parallel.

### 6.2.1 Evaluative Aims

We do not aim to perform an implementational comparison between vanilla BitTorrent and the proposed approaches, e.g., regarding protocol overhead and technical aspects to realize either approach. This is out of the scope of this paper. The goal of our evaluation is to shed light on the feasibility and potential of the two approaches based on the newly discovered observations from our studies.

For both approaches we wish to discover, *(i)* does the approach increase file availability in torrents with ordinarily unavailable seeders, and *(ii)* what are the implications of this in regard to download performance. We aim to investigate these factors on both a per-torrent and system-wide basis to explore how the effects of the approaches impact both perspectives on BitTorrent.

### 6.2.2 Input to the experiments

**Selecting the Torrents:** Our trace data encompasses tens of thousands of torrents over a period of several weeks, far more then the simulator is able to handle. Hence, we chose a random subset of 100 torrents from the set of torrents affected by seedless states with varying file sizes between 3-1500 MB and a per-torrent monitoring period of at least four weeks [5]. The logs of these torrents contains data of more than 235,000 downloads.

**User behavior:** To model the access pattern of torrents, we do not use any artificial peer arrival function. Instead, we bring up new peers as well as reoccurring seeders according to the trace logs. To model the number of swarms that a peer joins we calculate the probability distribution over our entire data set. Any user that cannot download the file within 36 hours aborts the download [6]. Finally, after finishing their downloads, users stay online as seeders based on the measurements from our microscopic crawlings (cf. Fig. 8).

**Speed distributions:** To have a representative bandwidth distribution, we first associate each IP address with a country, using a freely available geolocation database [12]. Based on the country of origin, the Ookla database [16] provides us with the median down/uplink capacity of each user [7].

**Failures in contribution histories:** To represent information inconsistencies in the distribution of contribution histories (e.g. due to churn), when encountering a new user in the cross-torrent approach, the contribution history is only known with a probability of 0.9. This represents a worst-case scenario, as the literature has reported a superior accuracy of 0.96% [17].

---

[5] We have also experimented with higher/smaller amount of torrents. Due to space constraints, we opt for presenting only a representative sample.

[6] We find through simulations that 36 hours is enough time to get a download success ratio over 99% in the presence of seeders for all access links and file sizes used in our experiments.

[7] We have also experimented with other datasets [9, 4] and obtained similar results.

| Protocol | Avg. seeding time (in hours) | Metric $D_S$ (in KBps) | $F_S$ (in %) |
|---|---|---|---|
| BT: Vanilla | 3.44 | 137.84 | 20.25 |
| ST: 2x seeding | 6.88 | 158.11 | 13.65 |
| ST: 5x seeding | 17.20 | 179.41 | 4.39 |
| ST: 10x seeding | 34.40 | 190.81 | 0.66 |
| CT: Persistent history | 3.44 | 138.75 | 0.13 |

**Table 2. Overview about system-level results.**

#### 6.2.3 Performance metrics

We utilise two performance metrics to evaluate the effectiveness of the approaches. The first is the average *downloading rate* of successful users ($D$) and the second is the fraction of *download abortions* ($F$). Both metrics are calculated on a per-torrent ($D_T, F_T$) and system-wide basis ($D_S, F_S$).

### 6.3 Comparative Results

Table 2 gives an overview of the three variants in which the measured seeding times are lengthened by a factor of either 2, 5, or 10. For comparability reasons, we assumed for the cross-torrent approach that users remain online after downloading as long as they stay in vanilla BitTorrent. The results presented in this table summarize the fraction of download abortions ($F_S$) and the average downloading rate ($D_S$) on a system-wide level. Some points worth noting:

- In the chosen set of torrents, 20% of downloads were not successful in vanilla BitTorrent.

- To maintain persistent file availability in the single-torrent approach, e.g. to ensure a system wide success rate for downloads $> 99\%$, the users must stay in average 10 times longer after downloading. This induces an average seeding time of more than 34 hours.

- The cross-torrent approach achieves a similar downloading failure ratio as the single-torrent variant lengthening the seeding times by a factor of 10. However, this is achieved without having to increase the seeding times beyond that currently observed in BitTorrent. With regard to download rates, the cross-torrent approach also performs similar to vanilla BitTorrent.

In addition to this table, Fig. 10 and Fig. 11 separately plot the fraction of aborted downloads and the average downloading rate, respectively, on a torrent basis. It can be observed that, in a few torrents, even a 10 times multiplication of seeding times only has a marginal effect on reducing the abortion ratio. On the other side, the cross-torrent approach obviously benefits from the available file replicas as the abortion ratio never exceeds the 2% threshold.

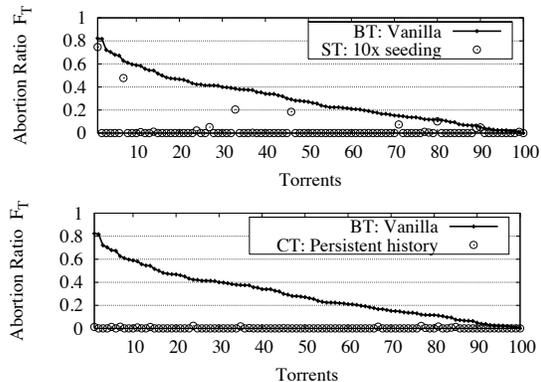

**Figure 10. Overview about the per-torrent abortion ratios.**

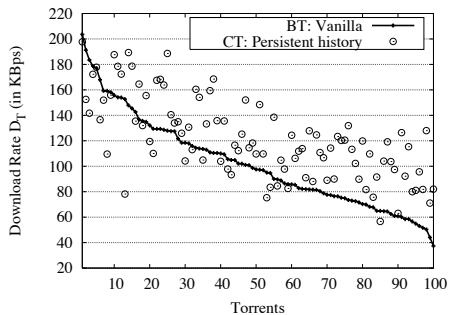

**Figure 11. Average download rates on a per-torrent basis.**

When examining the download performance on a per-torrent basis in Fig. 11, it can be observed that users applying the cross-torrent solution actually increase their download rate when compared to vanilla BitTorrent. For instance, the average downloading rate ($\bar{D}_T$) over all torrents is 102.26 KBps for vanilla BitTorrent and 121.65 KBps for the cross-torrent variant. As this is an average, it does not highlight the disparities between different torrents' performance based on popularity. Whilst not shown in the plots, the cross-torrent approach reallocates upload capacity from particularly popular torrents to other torrents. Therefore, 43.55% of the users that finish in vanilla BitTorrent gain a performance increase of 88% when applying cross-torrent collaboration. However, the remaining 56.44% of these users suffer from an average performance decrease of 22%.

### 6.4 Summary

To conclude, even when considering the seeding discontinuity of users, the single-torrent approach emerges as highly impracticable: to ensure a system-wide file availability of >99%, average seeding times of more than 34 hours are required. In contrast, the cross-torrent approach using persistent histories achieves this level of file availability easily. It therefore allows the 20% of users that could not complete their downloads in vanilla BitTorrent to effectively download the file. The download performance of the cross-torrent approach is on a system level equivalent to vanilla BitTorrent, and even improves download times on a per-torrent basis. Although this finding is at first glance surprising, the cross-torrent approach benefits from the significant increase in nodes (>20%) which now find content available. This, in turn, allows users that previously could not access the content to download at high rates; this compensates for the inherent performance disadvantages of peer selection policies that are optimised for fairness [6].

However, it must also be noted that the increase of file availability due to cross-torrent collaboration is achieved by a notable trade-off. That is, the download rate of more than half of the users that finish downloads with vanilla BitTorrent accounting degrades by 22%. This can be contrasted with a $88\%$ improvement for the remainder of peers.

## 7 Conclusions

This paper has investigated BitTorrent's unavailability problem in the wild and explored the feasibility of the potential solution-space. To achieve this, two large-scale measurements studies were performed to ascertain the characteristics, causes and repercussions of file unavailability in BitTorrent. Based on this, we made a number of interesting findings that offer the most accurate study of file availability in BitTorrent so far. Most notably, it was found that $(i)$ a lack of seeders often results in unavailability but *not* always, $(ii)$ the churn level, the fast replication of rare chunks and the population size largely defines a swarm's ability to survive without a seeder $(iii)$ unavailability usually occurs in cyclic periods with intermittent availability, and $(iv)$ unavailability often results in a chain effect that leads to future download failures.

Due to these new findings, the solution-space was also investigated to see how they affect both single and cross torrent solutions. It was found that the continuance of BitTorrent's single-torrent mechanisms can only address the problem with a 10 fold increase in seeding times. In contrast, great potential has been found in using the cross-torrent approach which maintains current performance levels whilst also achieving over 99% availability.